# Overcoming the Loss Conditioning Bottleneck in Optimization-Based PDE Solvers: A Novel Well-Conditioned Loss Function

Wenbo Cao[a,b,c], Weiwei Zhang[a,b,c],*

[a] *School of Aeronautics, Northwestern Polytechnical University, Xi'an 710072, China*
[b] *International Joint Institute of Artificial Intelligence on Fluid Mechanics, Northwestern Polytechnical University, Xi'an, 710072, China*
[c] *National Key Laboratory of Aircraft Configuration Design, Xi'an 710072, China*
* *Corresponding author. E-mail address: aeroelastic@nwpu.edu.cn*

**Abstract.** Optimization-based PDE solvers that minimize scalar loss functions have gained increasing attention in recent years. These methods either define the loss directly over discrete variables, as in Optimizing a Discrete Loss (ODIL), or indirectly through a neural network surrogate, as in Physics-Informed Neural Networks (PINNs). However, despite their promise, such methods often converge much more slowly than classical iterative solvers and are commonly regarded as inefficient. This work provides a theoretical insight, attributing the inefficiency to the use of the mean squared error (MSE) loss, which implicitly forms the normal equations, squares the condition number, and severely impairs optimization. To address this, we propose a novel Stabilized Gradient Residual (SGR) loss. By tuning a weight parameter, it flexibly modulates the condition number between the original system and its normal equations, while reducing to the MSE loss in the limiting case. We systematically benchmark the convergence behavior and optimization stability of the SGR loss within both the ODIL framework and PINNs—employing either numerical or automatic differentiation—and compare its performance against classical iterative solvers. Numerical experiments on a range of benchmark problems demonstrate that, within the ODIL framework, the proposed SGR loss achieves orders-of-magnitude faster convergence than the MSE loss. It retains the computational advantages of explicit schemes while attaining convergence efficiencies comparable to classical implicit solvers, offering new insights for developing advanced iterative schemes. Further validation within the PINNs framework shows that, despite the high nonlinearity of neural networks, SGR consistently outperforms the MSE loss. These theoretical and empirical findings help bridge the performance gap between classical iterative solvers and optimization-based solvers, highlighting the central role of loss conditioning, and provide key insights for the design of more efficient PDE solvers. All code and data used in this study are available at https://github.com/Cao-WenBo/StabilizedGradientResidual.

# 1 Introduction

The solution of partial differential equations (PDEs) is fundamental to modeling, simulating, and optimizing complex systems across science and engineering. Traditional numerical methods—based on finite difference, finite volume, or finite element discretization—have long served as the cornerstone of PDE solvers. These methods typically rely on iterative solvers, including classical stationary iterative methods, Krylov subspace methods, multigrid techniques, and others. Well-established and highly efficient for large-scale problems, they also offer key advantages such as numerical stability, conservation, and robustness. However, these methods face significant challenges when applied to high-dimensional or parametric problems, where the curse of dimensionality renders mesh-based discretization and solution computationally prohibitive [1, 2]. Furthermore, in PDE-constrained optimization tasks, classical solvers must be invoked repeatedly within outer optimization loops, leading to substantial computational overhead [3]. In inverse problems involving incomplete or noisy boundary data—such as inferring velocity and pressure fields from sparse concentration measurements—traditional solvers often become ineffective due to ill-posedness and insufficient boundary information [4].

In recent years, neural network solvers, such as physics-informed neural networks (PINNs) [5], the deep Ritz method [2], and the deep Galerkin method [1], have significantly advanced the way PDEs are solved. These methods work by optimizing neural network parameters to minimize the residuals of the PDEs while simultaneously enforcing boundary conditions and incorporating data constraints when available, thus enabling the direct approximation of PDE solutions. A growing body of PINN-inspired methods has since emerged, demonstrating remarkable success across a wide spectrum of problems in computational science and engineering [4, 6-9]. One notable advantage of these solvers is their natural ability to handle inverse problems: by combining sparse observations with physical constraints, they can reconstruct unobserved flow fields [10-12] or infer unknown parameters in the governing equations [5, 13-15]. Another key strength lies in their efficiency in solving parametric problems, a capability also known as operator learning. Rather than solving the PDE repeatedly for different conditions, neural network solvers can learn a unified mapping that directly predicts solutions under varying flow conditions, geometries, or initial conditions. As a result, they have become a promising tool for surrogate modeling [1, 3, 16-20]. These methods have also been extended to incorporate convolutional or graph convolutional networks, utilizing numerical differentiation for derivative computation [21-25].

Beyond neural network solvers, there also exist optimization-based approaches that solve PDEs by directly minimizing a loss function over the solution itself. One such method is known as optimizing a discrete loss (ODIL) [26], which formulates the PDE as an optimization problem, where the solution values on a discrete grid are treated as optimization variables. ODIL was originally proposed to efficiently solve inverse problems, bypassing the need for neural parameterization and offering faster efficiency and higher accuracy. This formulation provides greater flexibility in certain settings and demonstrates particular advantages when neural networks struggle to approximate the solution.

For clarity, we refer to these methods that solve PDEs by minimizing a loss function through optimization algorithms as optimization-based solvers (OBS), in contrast to classical iterative solvers (IS) that directly operate on the PDE residual. Despite their flexibility, generality, and wide applicability, both ODIL and PINNs often converge significantly slower than classical iterative solvers—even for well-conditioned problems—and may sometimes even fail to converge. This inefficiency is particularly puzzling in linear settings, where methods like the conjugate gradient (CG) or generalized minimal residual (GMRES) converge rapidly, while optimization-based methods often require orders of magnitude more iterations. In the context of PINNs, such inefficiencies have been extensively discussed. Krishnapriyan et al. [27] attribute common failure modes to the non-convex loss landscape, which is significantly more complex than that encountered in supervised learning. Wang et al. [28] used infinite-width NTK theory to propose that the subtle balance between the PDE residual and supervised components of the loss function could explain and possibly ameliorate training issues. Cao et al. [29] constructed controlled PDE systems with tunable Jacobian condition numbers to verify the correlation between the ill-conditioning of PINNs and that of the underlying PDE operator, and based on this insight, proposed time-stepping-oriented neural network (TSONN) to alleviate ill-conditioning. Rathore et al. [30] explained the optimization stagnation caused by residual loss in terms of the spectral density of the Hessian matrix of the loss function and improved PINNs performance by enhancing the gradient descent algorithm. De Ryck et al. [31] argued that the convergence rate of gradient descent in PINNs is related to the condition number of an operator, which is composed of the Hermitian square of differential operators of the PDE system. These different perspectives have inspired a variety of techniques to enhance the performance of PINNs.

In this work, we revisit this longstanding issue through simple and illustrative examples, aiming to provide a clear comparison between optimization-based solvers and classical iterative solvers. We analyze this discrepancy from the perspective of loss function design and system conditioning, and demonstrate that the commonly used mean squared error (MSE) loss squares the condition number, leading to significantly slower convergence—an insight consistent with De Ryck et al.'s analysis. Then, we propose an alternative loss function—Stabilized Gradient Residual (SGR)—that preserves the conditioning of the original PDE system. We provide a theoretical analysis of its properties and validate its performance in both ODIL and neural network solvers.

The remainder of this paper is organized as follows. Section 2 introduces the problem setup and provides theoretical insights of the differences between optimization-based and iterative solvers, highlighting the inefficiency of the MSE loss and proposing the SGR loss as an alternative. In Section 3, we validate the theoretical findings on a series of benchmark problems, demonstrating the effectiveness of the proposed SGR loss. Section 4 further interprets the SGR loss from the perspective of iterative update schemes and compares it with explicit iterative solvers. Finally, Section 5 concludes the paper and outlines potential directions for future research.

## 2 Loss Functions in Optimization-Based Solvers: Formulation and Conditioning

2.1 Problem setup

Many physical systems are governed by nonlinear partial differential equations (PDEs) of the form

$$\mathcal{N}[q](\boldsymbol{x}) = f_s(\boldsymbol{x}), \quad \boldsymbol{x} \in \Omega \tag{1}$$

where $\mathcal{N}$ denotes a nonlinear differential operator, $q(\boldsymbol{x})$ is the unknown solution, and $f_s(\boldsymbol{x})$ represents the source term. To solve such equations numerically, the domain $\Omega$ is discretized and the differential operator $\mathcal{N}$ is approximated using a numerical scheme—such as finite difference, finite volume, or finite element methods—resulting in a system of nonlinear algebraic equations. In many settings, it is natural to consider the linearized form of these equations, resulting in a linear system of the form

$$\boldsymbol{f} = A\boldsymbol{q} - \boldsymbol{b} = \boldsymbol{0} \tag{2}$$

where $A \in \mathbb{R}^{n \times n}$ is the Jacobian matrix (or a linearized residual operator), $\boldsymbol{q} \in \mathbb{R}^n$ is the discretized solution vector, and $\boldsymbol{b} \in \mathbb{R}^n$ is the source term evaluated at $\boldsymbol{q}^{(k)}$. This process is repeated iteratively to approach the solution of the original nonlinear problem.

There exist several approaches to solving such linear systems. Iterative solvers, such as the conjugate gradient (CG) method for symmetric systems and the generalized minimal residual (GMRES) method for nonsymmetric systems, directly operate on the residual $f$ and are widely used as baseline algorithms. Alternatively, the solution can be obtained by minimizing a scalar loss function—either directly defined over the discrete variable $q$, as in ODIL, or indirectly through the parameters of a neural network surrogate, as in PINNs with numerical differentiation (PINNs-ND). While these approaches differ in representation and implementation, they all involve the underlying operator $A$, and their convergence behavior is closely related to the condition number of $A$ or its transformations induced by the chosen loss function, which we examine in detail in the following subsection. In practice, PINNs employing automatic differentiation (PINNs-AD) are the more common setting, but they do not involve an explicit operator $A$; this distinction will be discussed later.

2.2 Loss functions for symmetric positive definite systems

We first consider the linear system resulting from the discretization of a symmetric positive definite (SPD) partial differential equation, such as the Poisson equation. In such cases, the discretization leads to a system matrix $A$ that is symmetric and positive definite. Here, we examine two distinct loss functions commonly used to solve the SPD system $f = Aq - b = 0$, and analyze their relationship to classical iterative solvers. In the context of machine learning, the mean squared error (MSE) is by far the most widely used objective for regression problems, including in neural network solvers such as PINNs. The MSE loss is defined as the squared $\ell^2$-norm of the residual vector, given by:

$$\mathcal{L}_{MSE} = \frac{1}{2}\|Aq - b\|_2^2 = \frac{1}{2}f^T f \tag{3}$$

Its gradient and Hessian are:

$$\nabla \mathcal{L}_{MSE} = A^T(Aq - b), \nabla^2 \mathcal{L}_{MSE} = A^T A \tag{4}$$

As a result, the conditioning of this optimization problem is governed by the spectrum of $A^T A$, leading to a condition number $\kappa(A^T A) = \kappa(A)^2$.

From classical optimization theory, solving the SPD system $f = Aq - b = 0$ is also equivalent to solving the following quadratic programming problem:

$$\mathcal{L}_{QP} = \frac{1}{2}q^T A q - q^T b \tag{5}$$

which we refer to as the quadratic programming (QP) loss. Its gradient is

$$\nabla \mathcal{L}_{QP} = A\boldsymbol{q} - \boldsymbol{b} \tag{6}$$

which is precisely the residual. The Hessian is $A$, so the optimization problem inherits the original condition number $\kappa(A)$, rather than its square. In fact, since the gradient exactly corresponds to the residual of the linear system, applying the CG method to minimize $\mathcal{L}_{QP}$ is mathematically equivalent to solving the system $A\boldsymbol{q} = \boldsymbol{b}$ using CG iterations.

Therefore, we observe that in the case of SPD systems, minimizing the QP loss results in the same condition number as the original linear system, suggesting that its convergence behavior closely aligns with that of classical iterative methods that directly operate on the residual. In contrast, the MSE loss amplifies the condition number by a factor of $\kappa(A)$, leading to significantly slower convergence. When these loss functions are used in PINNs-ND, the QP loss yields the following gradient and Hessian with respect to the neural network parameters $\boldsymbol{\theta}$:

$$\nabla_{\boldsymbol{\theta}} \mathcal{L}_{QP}(\boldsymbol{\theta}) = J^T A \boldsymbol{q} - J^T \boldsymbol{b}, \quad \nabla^2_{\boldsymbol{\theta}} \mathcal{L}_{QP}(\boldsymbol{\theta}) = J^T A J \tag{7}$$

while the MSE loss leads to:

$$\nabla_{\boldsymbol{\theta}} \mathcal{L}_{MSE}(\boldsymbol{\theta}) = J^T A^T A \boldsymbol{q} - J^T A^T \boldsymbol{b}, \quad \nabla^2_{\boldsymbol{\theta}} \mathcal{L}_{MSE}(\boldsymbol{\theta}) = J^T A^T A J \tag{8}$$

where $J$ denotes the Jacobian matrix of the network output $\boldsymbol{q}$ with respect to the network parameters $\boldsymbol{\theta}$. Therefore, this squared conditioning effect is expected to slow down the convergence of gradient-based optimization methods, even in the neural network setting. These theoretical insights will be substantiated through numerical experiments in Section 3.

2.3 Loss functions for nonsymmetric systems

For SPD systems, the QP loss consistently outperforms the MSE loss due to its favorable spectral properties. However, extending this formulation to more general nonsymmetric systems presents several theoretical difficulties. Its gradient and Hessian are

$$\nabla \mathcal{L}_{QP} = \frac{1}{2}(A + A^T)\boldsymbol{q} - \boldsymbol{b}, \nabla^2 \mathcal{L}_{QP} = \frac{1}{2}(A + A^T) \tag{9}$$

In the nonsymmetric case, this gradient introduces two major complications. First, since the Hessian is no longer symmetric, the loss function may not be convex, and in fact, may not be bounded from below—even in cases where the original matrix $A$ is positive definite. Second, even if the objective remains convex, the exact solution $\boldsymbol{q}_s$ to the system $A\bm{q} = \bm{b}$ does not correspond to a stationary point of the loss function, as the gradient does not vanish at the solution. To address this issue, we propose a new loss function, referred to as the gradient residual (GR) loss, defined as

$$\mathcal{L}_{GR} = \bm{f}^T(\bm{q} - \bm{q}_{\text{detach}}) + \frac{1}{2}\bm{f}_{\text{detach}}^T \bm{f}_{\text{detach}} \tag{10}$$

where the subscript "detach" indicates that the corresponding variable is treated as a constant and excluded from gradient computation. In other words, its value is used, but its derivative with respect to the optimization variables is zero. The first term of this loss is designed to provide a proper gradient with respect to $\bm{q}$, which precisely equals the residual $\bm{f}$, regardless of whether the system is symmetric or positive definite. Importantly, its value is intentionally constructed to be zero. The second term, equivalent to the value of the MSE loss, is included to track the actual loss magnitude. As a result, the loss always admits the true solution $\bm{q}_s$ as a stationary point.

Although the GR loss is formally defined as a loss function, it does not correspond to the gradient of any scalar function when the system matrix $A$ is nonsymmetric. This fact follows directly from Poincaré's lemma, which states that a vector field is a gradient field if and only if its curl vanishes—equivalently, $A$ must be symmetric for $F(\bm{q}) = A\bm{q} - \bm{b}$ to be a gradient. Consequently, the GR loss should not be interpreted strictly as an optimization objective. Instead, gradient-based methods with a learning rate $\eta$ applied to the loss induce an iterative scheme, referred to as the GR iteration, of the form

$$\bm{q}^{(k+1)} = \bm{q}^{(k)} - \eta(A\bm{q}^{(k)} - \bm{b}) \tag{11}$$

which can be understood as a numerical iteration analogous to classical stationary iterative methods. Under this interpretation, we also can view MSE as giving rise to distinct iterative update schemes of the form

$$\bm{q}^{(k+1)} = \bm{q}^{(k)} - \eta A^T(A\bm{q}^{(k)} - \bm{b}) \tag{12}$$

which corresponds to gradient descent applied to the normal equation $A^T A \bm{q} = A^T \bm{b}$. It can be shown that, under sufficiently small learning rates, both iteration schemes are guaranteed to converge. Detailed proofs and convergence conditions are provided in Appendix A. However, for nonsymmetric systems, the GR iteration may exhibit noticeable oscillations or even instability. In contrast, the MSE iteration, although significantly slower due to the squared condition number introduced by $A^T A$, benefits from the fact that $A^T A$ is always symmetric and positive definite, leading to more stable convergence behavior. Motivated by these complementary properties, we construct a new loss function, referred to as the stabilized gradient residual (SGR) loss, defined as

$$\mathcal{L}_{SGR} = (1-\omega)\boldsymbol{f}^T\boldsymbol{f} + \omega\boldsymbol{f}^T(\boldsymbol{q}-\boldsymbol{q}_{\text{detach}}) + \omega\boldsymbol{f}_{\text{detach}}^T\boldsymbol{f}_{\text{detach}} \qquad (13)$$

It can be observed that the SGR loss yields exactly the same value as the MSE loss, while its gradient corresponds to a weighted combination of the GR and MSE gradients. The purpose of SGR is to stabilize the GR loss using the MSE loss in order to suppress severe oscillations. As previously mentioned, strictly speaking, Equation (13) corresponds to an iterative update scheme rather than a scalar loss function. However, in conventional iterative solvers, computing $A^T\boldsymbol{f}$ is typically difficult and computationally expensive. In contrast, under the framework of optimization-based solvers, $A^T\boldsymbol{f}$ naturally emerges from the backpropagation of the loss, incurring no additional computational cost. Furthermore, it is worth noting that the SGR loss completely avoids the explicit use of $A$ and $\boldsymbol{b}$, making it readily extensible to both PINNs-ND and PINNs-AD frameworks.

## 3 Results

This section presents numerical experiments that validate the theoretical insights developed in Section 2. In particular, we compare the convergence behavior, sensitivity to conditioning, and optimization stability of different loss formulations—namely MSE, QP, GR, and SGR—as well as classical iterative solvers, including CG and GMRES, when applied to PDE systems arising from both symmetric and nonsymmetric PDE discretization. We also demonstrate the extensibility of these loss functions to neural network solvers. The evaluation aims to reveal how different loss formulations and problem characteristics jointly affect the convergence and stability of various solvers.

In all neural network–based experiments, the neural network architecture consists of a fully connected network with 4 hidden layers, each containing 128 neurons and using the tanh activation function. The Adam optimizer is employed with a fixed learning rate of 0.0002. While this setting may result in slower convergence, it enables a fair comparison among different methods and is sufficient to achieve acceptable solution accuracy. The collocation points are selected to coincide with the grid points used in the finite difference discretization of each specific problem.

3.1 2D Poisson equation

We begin with the solution of a standard two-dimensional Poisson equation of the

form

$$q_{xx} + q_{yy} = f_s(x,y), (x,y) \in \Omega = [0,1]^2 \qquad (14)$$

where the source term $f_s(x,y)$ and Dirichlet boundary conditions are chosen such that the exact solution is $q(x,y) = \sin(\pi(2x)^2)\sin(\pi y)$. This benchmark problem has also been used in prior works for validating ODIL [26]. To construct symmetric positive definite systems with varying condition numbers, we discretize the equation using second-order central differences on uniform grids of increasing resolution: $n = 50 \times 50, 100 \times 100, 200 \times 200, 400 \times 400$. The resulting matrices $A$ are sparse, symmetric, and positive definite. The corresponding condition numbers are reported in Table 1.

Table 1. Condition numbers of the PDE system across different grid resolutions.

| $n$ | $50^2$ | $100^2$ | $200^2$ | $400^2$ |
|---|---|---|---|---|
| $\kappa(A)$ | 972 | 3971 | 16056 | 64618 |

We first solve each discretized system $A\boldsymbol{q} = \boldsymbol{b}$ using the CG method and the GMRES method. In addition, we solve the corresponding normal equations $A^T A \boldsymbol{q} = A^T \boldsymbol{b}$, which arise from minimizing the MSE loss. GMRES is restarted every 300 iterations. The results are shown in Figure 1. The error plotted represents the relative $L_2$ norm of the solution vector, computed as $\|\boldsymbol{q} - \boldsymbol{q}_{ref}\|_2 / \|\boldsymbol{q}_{ref}\|_2$. As the grid resolution increases—resulting in larger system sizes and higher condition numbers—we observe that both CG and GMRES converge more slowly. Furthermore, the normal equations exhibit significantly slower convergence compared to the original system, clearly highlighting the influence of the condition number on iterative solvers and confirming the squared amplification effect induced by $A^T A$. Interestingly, for the normal equations, CG consistently converges much faster than GMRES. This is because the system is SPD, which allows CG to fully exploit its optimal convergence properties, while GMRES, which does not benefit from symmetry, performs less efficiently. Nevertheless, to ensure general applicability to PDE systems that give rise to nonsymmetric operators, it remains essential to evaluate the performance of GMRES as well.

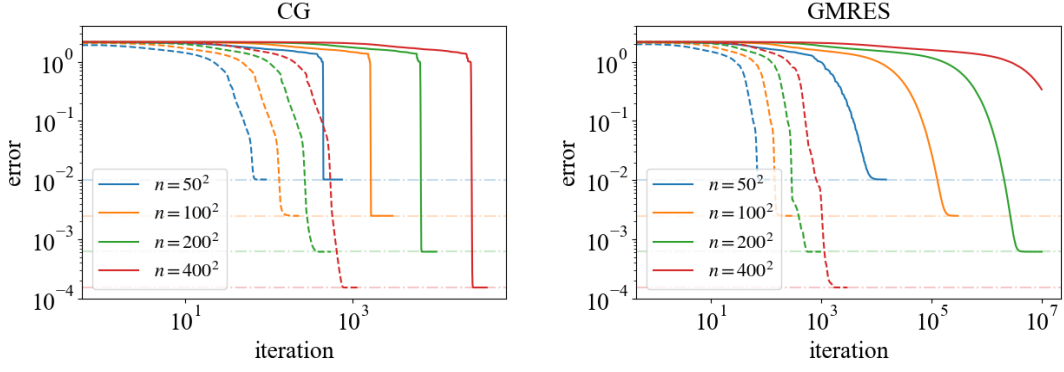

Figure 1. Convergence histories of iterative solvers on Poisson systems. Different colors indicate different grid resolutions. Horizontal dashed lines represent the final error level after full convergence for each grid. Dashed curves correspond to solvers applied to the original system, while solid curves correspond to the normal equations.

Next, we solve the Poisson system using optimization-based solvers, specifically the ODIL method developed for solving discrete PDE systems via optimization. Two popular optimizers, limited-memory Broyden-Fletcher-Goldfarb-Shanno (LBFGS) and Adam, are employed to minimize two distinct loss functions: the MSE loss and the QP loss. The Adam optimizer is used with a fixed learning rate of 0.001. The corresponding convergence histories are presented in Figure 2. Similar to iterative solvers, we also observe that the convergence slows down as the grid is refined and the condition number increases. Since the MSE loss corresponds to the normal equations, its convergence is significantly slower than that of the QP loss, consistent with the observations in Figure 1. Due to its second-order convergence properties, LBFGS achieves faster convergence under the MSE loss; however, excessive ill-conditioning leads to difficulties in approximating the Hessian matrix accurately, ultimately degrading convergence accuracy. In Figure 2, Adam demonstrates relatively slow convergence with the QP loss. We emphasize that this behavior is closely related to the choice of learning rate. To illustrate this, we further examine the convergence histories of Adam under different learning rates, as shown in Figure 3. We observe that a larger learning rate can further accelerate the convergence of the QP loss, whereas the MSE loss shows little to no improvement in convergence speed. However, increasing the learning rate also leads to greater oscillations and higher loss values. This issue can be mitigated by employing an appropriate learning rate decay strategy, which will be further demonstrated in Section 3.2.

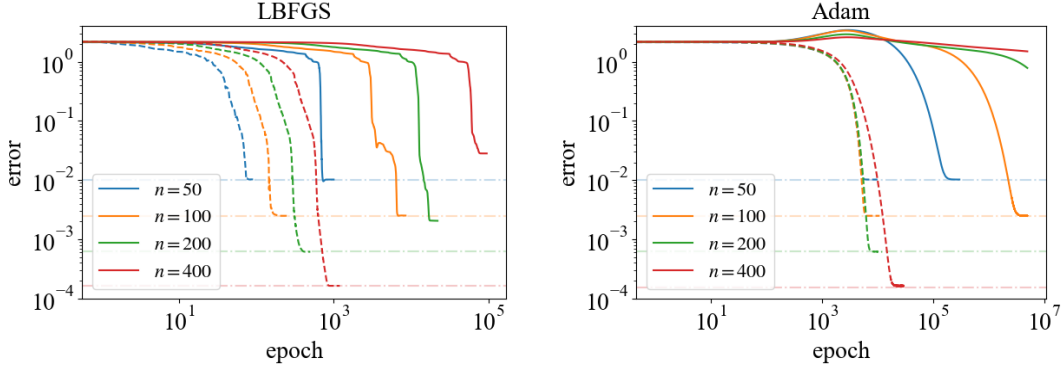

Figure 2. Convergence histories of ODIL on Poisson systems. Dashed curves correspond to results obtained using the QP loss, while solid curves represent those using the MSE loss.

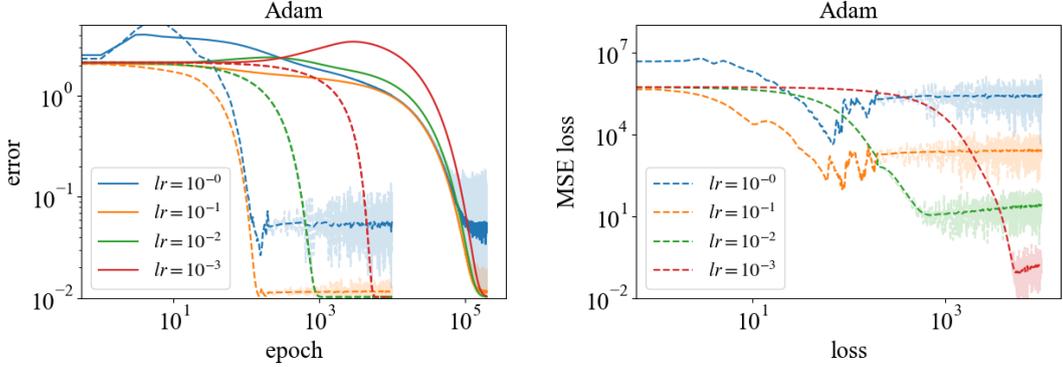

Figure 3. Convergence histories of ODIL on Poisson systems using Adam optimizer with different learning rate. The right figure shows the convergence curve of the QP loss evaluated using the $\ell_2$ norm of the residual, i.e., the MSE loss. The transparent lines represent the actual convergence curves, while the opaque lines represent Gaussian-filtered convergence curves for clearer visualization.

Next, we further evaluate the performance of the proposed loss formulations in the context of PINNs, using both PINNs-ND and PINNs-AD. It is important to note that in the PINNs-AD setting, the system matrix $A$ and vector $b$ are not explicitly available. Therefore, we adopt the GR loss as a surrogate for the QP loss, which is known to degenerate to the QP loss in the case of SPD systems. In addition, for PINNs-ND, the boundary conditions are already embedded into the solution during the numerical discretization scheme. For PINNs-AD, to eliminate the influence of relative weighting between the PDE residual and boundary condition loss terms, the boundary conditions are imposed as hard constraints. Specifically, the predicted solution is reparametrized as $\hat{q} = 100x(1-x)y(1-y)q_{NN}(x,t;\boldsymbol{\theta})$.

The convergence histories are shown in Figure 4. In the PINNs-ND setting, we observe a clear trend of slower convergence as the condition number of the underlying linear system increases. Moreover, the QP loss consistently outperforms the MSE loss across all grid resolutions, consistent with the results observed in both classical iterative solvers and optimization-based methods without neural networks. However, when embedded in the neural network framework, both loss functions exhibit more pronounced oscillations during training. This behavior is attributed to the inherent nonlinearity and nonconvexity of neural networks, which can introduce additional instability in the optimization dynamics. In particular, the LBFGS optimizer becomes unstable and diverges when applied to highly ill-conditioned systems, highlighting its sensitivity to curvature in complex, nonconvex loss landscapes.

All the above experiments validate the theoretical insights presented in Section 2.2, namely that the condition number of the system significantly impacts convergence behavior, and that the quadratic amplification of the condition number induced by the MSE loss leads to slower convergence. In contrast, for PINNs-AD, where the system matrix $A$ derived from numerical discretization is not explicitly involved, we do not observe a clear correlation between convergence speed and the number of grid points (i.e., collocation points). Moreover, no significant difference in convergence behavior is observed between the MSE and QP losses in the PINNs-AD context. However, this does not imply that the quadratic amplification of the condition number by the MSE loss is absent in general; rather, it likely reflects the relatively mild conditioning of this particular test case. This will be further confirmed by the numerical examples in Sections 3.2 and 3.3.

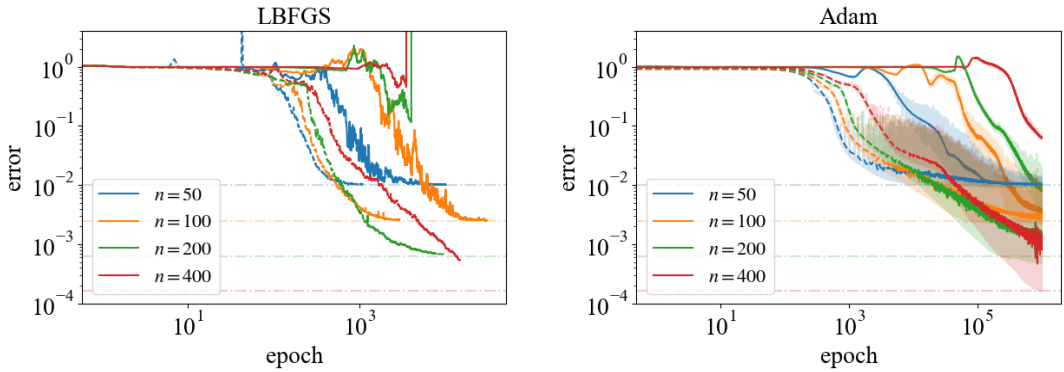

Figure 4. Convergence histories of PINNs-ND on Poisson systems. Dashed curves correspond to results obtained using the QP loss, while solid curves represent those using the MSE loss.

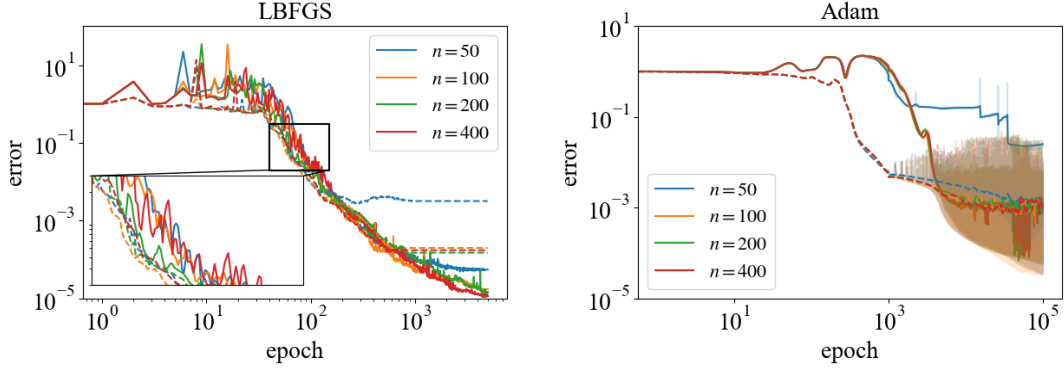

Figure 5. Convergence histories of PINNs-AD on Poisson systems. Dashed curves correspond to results obtained using the QP loss, while solid curves represent those using the MSE loss.

3.2 Allen-Cahn equation

In this subsection, we further investigate a nonlinear, non-symmetric, and time-dependent PDE with sharp interface dynamics—the Allen–Cahn equation—which serves as a standard benchmark for PINNs. This example is designed to validate the theoretical insights discussed in Section 2.3. The Allen–Cahn equation is given by:

$$\frac{\partial q}{\partial t} - 0.0001\frac{\partial^2 q}{\partial x^2} + 5q^3 - 5q = 0, x \in [-1,1], t \in [0,1] \tag{15}$$
$$q(x,0) = x^2 \cos(\pi x)$$

Periodic boundary conditions are imposed in space.

To discretize the system, we employ a second-order central difference scheme in space and a first-order forward Euler scheme in time, on uniform grids with $n = n_x \times n_t = 257 \times 101$ points. The problem is formulated in a space–time unified manner, consistent with the PINNs paradigm, which treats both space and time as inputs to the neural network solver. Due to the nonlinear of the PDE, the system must be repeatedly linearized throughout the solution process. At each iteration, we solve the resulting linearized system—or its normal equations—using non-restarted GMRES. We refer to this procedure as the inner iteration. We consider different values of the inner iteration count, denoted as *n_linear*, which specifies the maximum number of GMRES steps used to solve each linearized system. A large *n_linear* may yield more accurate inner solutions but incurs higher computational cost per outer iteration, while a small *n_linear* reduces the inner cost but may require significantly more outer iterations to reach convergence. By varying *n_linear*, we aim to identify the optimal trade-off between inner solver accuracy and overall computational efficiency. The results in Figure 6 show that when solving the normal equations in the inner loop, significantly more inner iterations are required. This is due to the substantial amplification of the condition number caused by the normal equations, ultimately leading to several orders of magnitude more iterations compared to solving the original system.

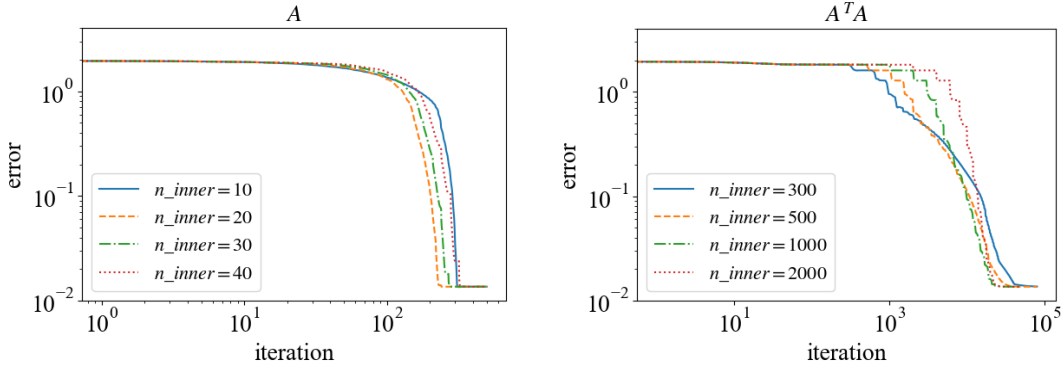

Figure 6. Convergence histories of iterative solvers on Allen-Cahn equation.

Next, we solve the same PDE system using the SGR loss within the ODIL framework, varying the weighting parameter $\omega$. Depending on its value, the SGR loss degenerates to the GR loss when $\omega=1$, and to the MSE loss when $\omega=0$. Within the ODIL framework, the problem is formulated and solved directly as a nonlinear optimization problem, without the need for repeated local linearization as required by iterative solvers. As shown in Figure 7, both the LBFGS and Adam optimizers exhibit the slowest convergence when $\omega=0$, while increasing $\omega$ generally accelerates convergence. However, when $\omega$ becomes too large, the LBFGS optimizer suffers from divergence. This can be attributed to the fact that LBFGS constructs an approximate symmetric Hessian using historical gradients, while the actual Hessian of the SGR loss is non-symmetric unless $\omega=0$. This structural mismatch may lead to unpredictable and unstable convergence behavior. In contrast, as a first-order method, Adam exhibits stable convergence for small $\omega$, albeit slowly, while larger $\omega$ values lead to faster convergence but also introduce noticeable oscillations. Figure 8 further shows that using larger learning rates accelerates convergence but increases the final residual. Applying a learning rate decay strategy achieves both rapid and stable convergence, with final losses approaching machine precision. These results suggest that within the ODIL framework, the SGR loss combined with the Adam optimizer offers a robust and efficient solution strategy. With properly chosen hyperparameters, it can approach the convergence efficiency of fully implicit iterative methods.

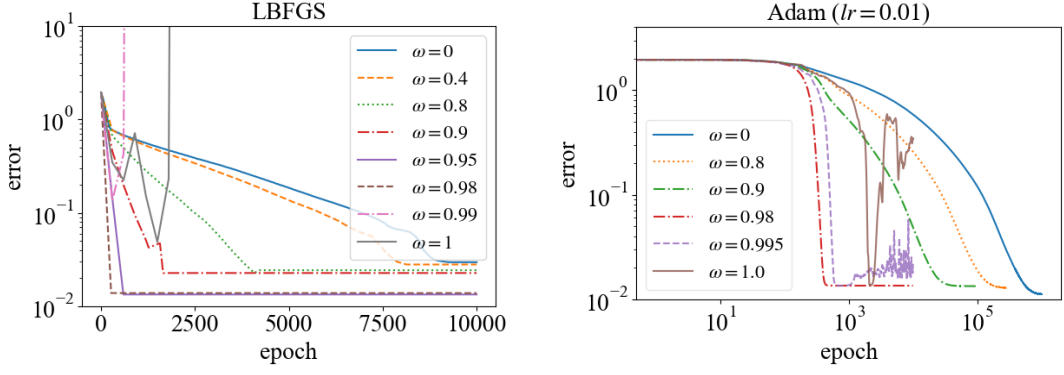

Figure 7. Convergence histories of ODIL using the SGR loss on the Allen–Cahn equation.

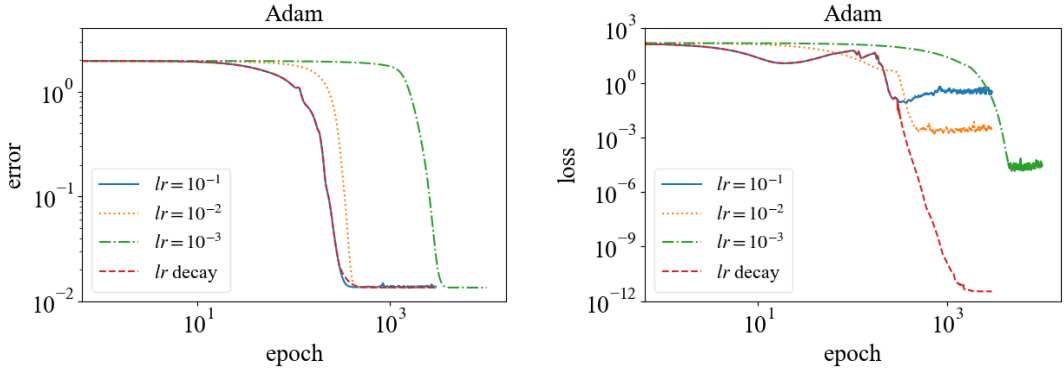

Figure 8. Convergence histories of ODIL with the Adam optimizer under different learning rate strategies on the Allen–Cahn equation. A fixed value of $\omega = 0.98$ is used in all cases. The figure compares multiple constant learning rates with a decaying schedule ("$lr$ decay") that starts at 0.1 and decays by a factor of 0.5 every 300 steps, with a minimum value of $10^{-5}$.

Next, we further evaluate the performance of the SGR loss in the context of PINNs, using both PINNs-ND and PINNs-AD. To facilitate the implementation of the SGR loss in PINNs-AD, the loss is reformulated as the $\ell^2$-norm of the residual vector of the underlying PDE system, which incorporates the PDE residual, boundary condition residual, and initial condition residual. Following the approach in [29], it is expressed as:

$$f(q) = \begin{bmatrix} \lambda_{PDE}\, g(q) \\ \lambda_{BC} \sqrt{N_g / N_h}\ h(q) \\ \lambda_{IC} \sqrt{N_g / N_i}\ i(q) \end{bmatrix} = 0 \qquad (16)$$

where $g(q)$, $h(q)$, and $i(q)$ denote the PDE residual vector, boundary condition residual vector, and initial condition residual vector, respectively; and $N_g$, $N_h$, and $N_i$ represent their corresponding dimensions. We use the relative weights $\lambda_{PDE}=1, \lambda_{BC}=2, \lambda_{IC}=5$. The results shown in Figure 9 indicate that although the advantage of SGR over MSE is not as significant as observed in the ODIL framework, it still provides noticeable improvements. Moreover, the convergence generally becomes faster as $\omega$ increases, but this comes at the cost of increased oscillations, which can be attributed to the inherent nonlinearity and nonconvexity of neural networks. These findings also suggest that, even in PINNs-AD—where the loss does not involve an explicit appearance of the discretized operator $A$—the underlying influence of $A$ conditioning persists. This empirically validates the theoretical insight that the MSE loss implicitly amplifies the condition number through the normal equations, and that the SGR loss can effectively mitigate this ill-conditioning. The pronounced oscillations observed in the convergence curves suggest a possible direction for future work, such as incorporating adaptive $\omega$ scheduling or learning rate decay to stabilize training dynamics. More importantly, a key challenge lies in improving the network architecture or training strategy within the neural network context to support larger values of $\omega$, thereby achieving convergence rates closer to those of classical iterative solvers.

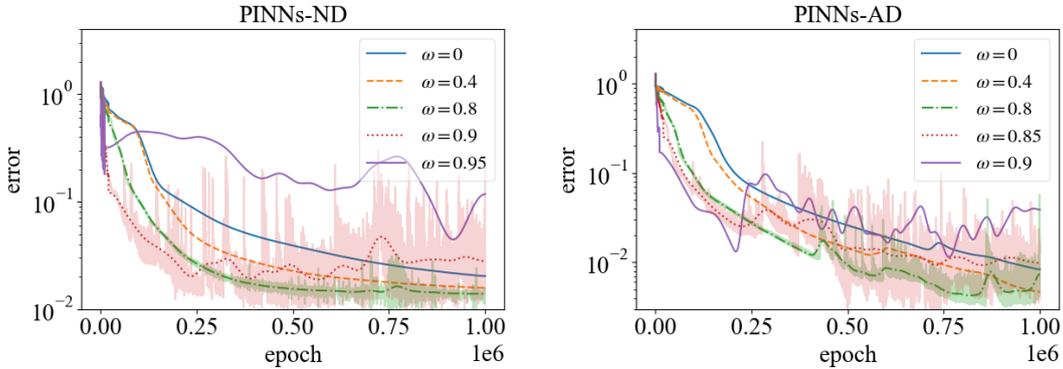

Figure 9. Convergence histories of PINNs-ND and PINNs-AD on the Allen–Cahn equation. Transparent lines indicate the raw convergence trajectories (partially shown for clarity), while opaque lines represent Gaussian-filtered curves for improved visual clarity.

### 3.3 Navier-Stokes equations

Finally, we consider the lid-driven cavity flow governed by the incompressible Navier–Stokes equations—a classical benchmark problem in both CFD and PINNs literature. The governing equations and boundary conditions are given by:

$$\begin{aligned}
&\boldsymbol{u}\cdot\nabla\boldsymbol{u}+\nabla p-\Delta\boldsymbol{u}/\mathrm{Re}=0\\
&\nabla\cdot\boldsymbol{u}=0\\
&\boldsymbol{u}=(1,0)\quad\text{in }\Gamma_0\\
&\boldsymbol{u}=(0,0)\quad\text{in }\Gamma_1
\end{aligned} \qquad (17)$$

where $\boldsymbol{u}=(u,v)$ is velocity vector, $p$ is pressure. The computational domain $\Omega=(0,1)\times(0,1)$ is a two-dimensional square cavity, where $\Gamma_0$ is its top boundary and $\Gamma_1$ is the other three sides. Despite its simple geometry, the driven cavity flow retains a rich fluid flow physics manifested by multiple counter rotating recirculating regions on the corners of the cavity as *Re* increases [32].

To discretize the system, we employ a second-order central difference scheme on uniform grids with $n=200\times200$ points in total. We first solve the problem with Reynolds number $Re=500$, using the SGR loss within the ODIL framework while varying the weighting parameter $\omega$. The results are shown in the left panel of Figure 10. We observe that the MSE loss (i.e., $\omega=0$) leads to extremely slow convergence, failing to reach an acceptable error threshold within a reasonable computational budget. As $\omega$ increases, convergence becomes significantly faster; however, when $\omega$ is too large, divergence is observed, consistent with earlier findings. Additionally, for small values of $\omega$, the final solution exhibits reduced accuracy, likely due to the highly ill-conditioned system. This issue can potentially be mitigated by applying a learning rate decay strategy. To further validate this, we consider a more challenging flow condition with $Re=2500$. As shown in the right panel of Figure 10, we find that larger learning rates lead to faster initial convergence but compromise final accuracy, while smaller learning rates yield more accurate results but converge more slowly. In both cases, the loss function tends to plateau at a relatively large value. In contrast, employing a learning rate decay schedule ensures a synchronized reduction in both the error and the residual, enabling faster and more stable convergence.

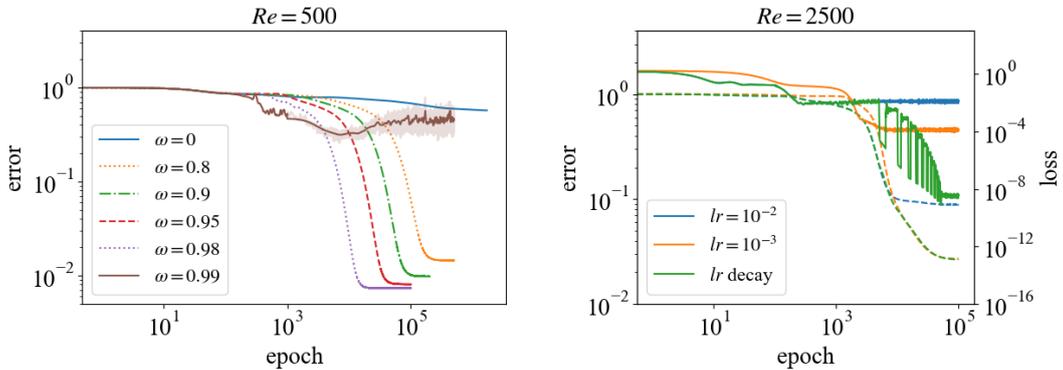

Figure 10. Convergence histories of ODIL using the SGR loss on the lid-driven cavity flow problem. The left panel uses a fixed learning rate ($lr = 0.01$) with varying values of $\omega$, while the right panel compares different learning rate strategies with fixed values of $\omega = 0.97$. In the right panel, dashed lines represent error convergence curves, and solid lines represent loss convergence curves. The "$lr$ decay" strategy starts at 0.01 and decays by a factor of 0.5 every 5000 steps, with a minimum learning rate of $10^{-5}$.

Next, we further evaluate the performance of the SGR loss in the context of PINNs. We consider the challenging flow condition with $Re = 2500$. We use the relative weights $\lambda_{PDE} = \lambda_{BC} = 1$. The results, presented in Figure 11, show that the SGR loss with a properly chosen $\omega$ significantly outperforms the MSE loss. In the case of PINNs-ND, the final relative error reaches 0.03, which is close to the optimal convergence accuracy of 0.026 achievable on the current grid resolution—demonstrating satisfactory performance. For PINNs-AD, although the final error remains relatively large, a substantial improvement over the MSE baseline is still observed. This performance gap highlights once again the potential for future work, such as incorporating adaptive $\omega$ strategies or learning rate decay mechanisms to further stabilize training dynamics and improve convergence in more challenging scenarios.

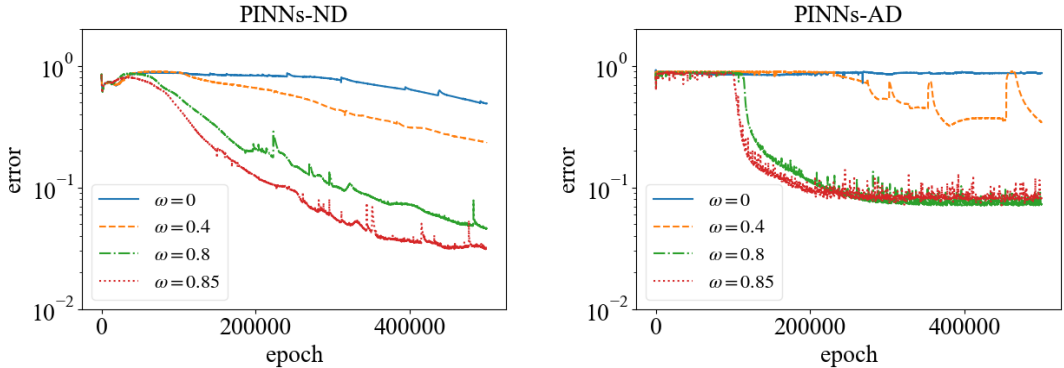

Figure 11. Convergence histories of PINNs-ND and PINNs-AD on the lid-driven cavity flow problem.

## 4 Discussions

We begin by reaffirming the fundamental connection between iterative solvers and optimization-based solvers without neural networks (i.e., ODIL). Both are, in essence, numerical methods based on stepwise updates of variables. The goal of an iterative solver is to drive the residual toward zero, while an optimization solver seeks to minimize the gradient of a loss function. When the gradient coincides with the residual, the two approaches become equivalent to some extent. For example, in the case of SPD systems, solving the linear system using the CG iteration is exactly equivalent to minimizing the QP loss using the CG method. In earlier sections, we introduced the SGR as a loss function and demonstrated its significant convergence advantages in optimization-based solvers. However, the significance of SGR extends beyond the optimization perspective. From the viewpoint of numerical iteration, the SGR scheme within the ODIL framework can in fact be interpreted as an advanced iterative method.

In particular, when the weighting parameter $\omega=1$, the SGR iteration reduces to the GR iteration defined in Equation (11), which precisely corresponds to the explicit time-marching scheme (Equation (18)) widely used in CFD. As is well known, time-marching methods are a core technique in CFD for solving both steady and unsteady systems, typically categorized into explicit and implicit schemes. Explicit methods are easy to implement, highly parallelizable, memory-efficient, and inexpensive per iteration. However, their convergence is usually restricted by the Courant–Friedrichs–Lewy (CFL) condition, which imposes strict limits on the time step size required for stability. In contrast, implicit schemes involve solving a linear system at each step. While they are more complex to implement, require higher memory usage, and are less amenable to parallelization, they achieve significantly fewer overall iterations and have thus become the dominant choice in large-scale engineering simulations.

$$\frac{\boldsymbol{q}^{(k+1)} - \boldsymbol{q}^{(k)}}{\Delta \tau} = -\boldsymbol{f}(\boldsymbol{q}^{(k)}) \tag{18}$$

In contrast, the SGR scheme does not fall strictly within the conventional categories of either explicit or implicit schemes. On one hand, it inherits many favorable properties of explicit methods—such as ease of implementation, parallelization-friendliness, and low per-step computational and memory costs. On the other hand, it significantly relaxes the stringent stability constraints typically imposed by the CFL condition, thereby allowing substantially larger step sizes (interpreted as learning rates in the optimization context). Figure 12 presents the convergence histories of explicit time marching (ETM) and SGR scheme for the Poisson and Navier–Stokes equations. For the Poisson case, a grid resolution of $n=100\times100$ is used, while for Navier–Stokes, a resolution of $n=200\times200$ is adopted. In both cases, ETM employs the empirically determined maximum stable time step, with $\Delta\tau=2\times10^{-5}$ for the Possion equation and $\Delta\tau=2\times10^{-4}$ for the Navier–Stokes equations. The results demonstrate that SGR requires one to two orders of magnitude fewer iterations to converge compared to ETS, while incurring less than twice the memory usage and computational time per iteration (see Table 2). These findings indicate that SGR strikes a favorable balance by retaining the computational efficiency and simplicity of explicit methods while significantly alleviating their stability limitations. This opens up promising directions for the development of next-generation iterative schemes. Furthermore, the physical meaning and stability mechanisms underlying SGR's ability to tolerate substantially larger time steps remain to be fully understood, presenting promising avenues for future theoretical research and the design of even more advanced iterative solvers.

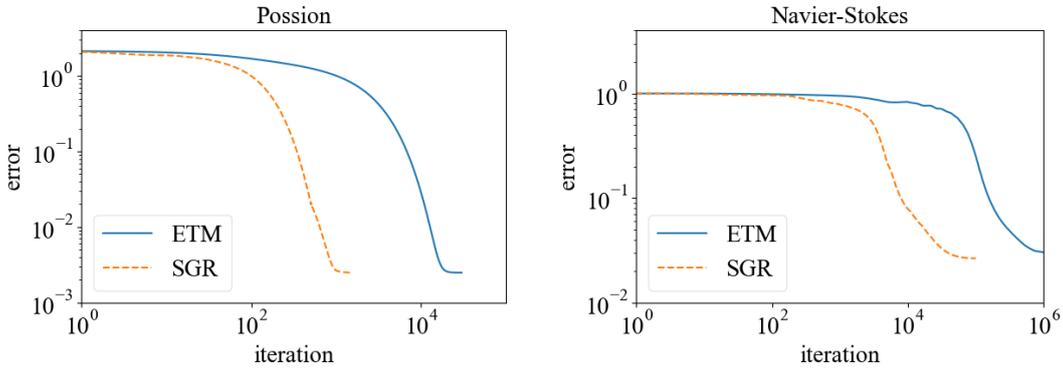

Figure 12. Convergence histories of explicit time marching (ETM) and SGR iteration. For the Poisson equation, the SGR iteration uses $\omega = 1$; for the Navier–Stokes equation, $\omega = 0.97$. For the Poisson equation, the learning rate starts at 0.1 and decays by 0.5 every 500 steps, down to a minimum of $10^{-5}$. For the Navier–Stokes equation, it starts at 0.01 and decays by 0.5 every 5000 steps, also with a minimum of $10^{-5}$.

Table 2. Comparison of memory usage and computational cost between explicit time marching and SGR iteration for different PDE systems (time reported per 1000 iterations).

| PDE system | Explicit time marching | | SGR iteration | |
| --- | --- | --- | --- | --- |
| | Memory (GB) | Time (s) | Memory (GB) | Time (s) |
| Poisson | 0.0084 | 3.9 | 0.0086 | 5.9 |
| Navier-Stokes | 0.0117 | 15.3 | 0.0147 | 23.1 |

## 5 Conclusions

This work systematically analyzes the inefficiency of optimization-based solvers for partial differential equations (PDEs), identifying its root cause. We show that the commonly used mean squared error (MSE) loss implicitly corresponds to solving the normal equations of the original discrete system, whose condition number is the square of that of the original problem. This squared conditioning severely amplifies the ill-conditioning and leads to slow convergence, even when classical iterative methods are used to solve the normal equations. This finding explains why optimization-based solvers, including PINNs and ODIL, exhibit much slower convergence compared to classical iterative solvers.

To address this issue, we propose the Gradient Residual (GR) loss that directly use the PDE residual as the gradient of the solution vector, thereby solving the original system rather than its normal equations. We further introduce a stabilized version, the Stabilized Gradient Residual (SGR) loss, which combines GR with MSE through a weighted sum. This approach effectively suppresses the oscillations observed with GR in nonsymmetric systems. Experimental results on various benchmark problems demonstrate that SGR loss can accelerate convergence by several orders of magnitude under the ODIL framework, approaching the efficiency of implicit iterative solvers. It also significantly improves convergence in the PINNs framework, showing strong generality and robustness.

Looking ahead, a central open question is how to improve neural network architectures and training strategies to enable the use of larger weighting parameters ($\omega$) in the SGR loss, along with appropriate adaptive scheduling for both $\omega$ and the learning rate to stabilize the training dynamics. Achieving this would significantly narrow the gap in convergence speed between optimization-based solvers and classical iterative solvers. Moreover, the SGR loss not only functions as an effective optimization objective, but also motivates a new class of iterative schemes, offering a fresh perspective for developing next-generation hybrid solvers that seamlessly combine classical numerical methods with modern optimization techniques.

## Data Availability Statement

The data that support the findings of this study are available from the corresponding author upon reasonable request.

## Conflict of Interest Statement

## Acknowledgments

We would like to acknowledge the support of the National Natural Science Foundation of China (No. 92152301).

## Appendix A

Given the physical system derived from the discretization of a PDE, the uniqueness of the solution ensures that the matrix $A$ is invertible, and the stability of the solution implies that $A$ is positive definite (though not necessarily symmetric). That is, all eigenvalues of $A$ have positive real parts. We aim to prove the convergence of the iterative scheme $\boldsymbol{q}^{(k+1)} = \boldsymbol{q}^{(k)} - \eta(A\boldsymbol{q}^{(k)} - \boldsymbol{b})$ to the unique solution $\boldsymbol{q}_s = A^{-1}\boldsymbol{b}$ under the appropriate choice of the step size $\eta > 0$.

Define the error vector at step $k$ as $\boldsymbol{e}^{(k+1)} = \boldsymbol{q}^{(k+1)} - \boldsymbol{q}_s$. The error propagation is

given by:

$$\boldsymbol{e}^{(k+1)} = \boldsymbol{q}^{(k+1)} - \boldsymbol{q}_s = [\boldsymbol{q}^{(k)} - \eta(A\boldsymbol{q}^{(k)} - \boldsymbol{b})] - \boldsymbol{q}_s \quad (A.1)$$

Substituting $A\boldsymbol{q}_s = \boldsymbol{b}$, we have:

$$A\boldsymbol{q}^{(k)} - \boldsymbol{b} = A\boldsymbol{q}^{(k)} - \boldsymbol{q}_s = A\boldsymbol{e}^{(k)} \quad (A.2)$$

Thus,

$$\boldsymbol{e}^{(k+1)} = \boldsymbol{e}^{(k)} - \eta A\boldsymbol{e}^{(k)} = (I - \eta A)\boldsymbol{e}^{(k)} = G\boldsymbol{e}^{(k)} \quad (A.3)$$

where $I$ is the identity matrix. The convergence of the iteration depends on the spectral radius of the iteration matrix $G$, denoted by $\rho(G)$. Specifically, the iteration converges to zero error for any initial error if and only if $\rho(G) < 1$.

Since $A$ is positive definite (but not necessarily symmetric), all eigenvalues $\lambda$ of $A$ have positive real parts, i.e., $\text{Re}(\lambda) > 0$. The eigenvalues of $G$ are $\mu = 1 - \eta\lambda$. We need to ensure that $|1 - \eta\lambda| < 1$ for every eigenvalue $\lambda$ of $A$.

Let $\lambda = a + bj$ with $a > 0$ and $j = \sqrt{-1}$. Then,

$$\mu = 1 - \eta(a + jb) = (1 - \eta a) - j\eta b \quad (A.4)$$

The squared modulus of $\mu$ is:

$$|\mu| = (1 - \eta a)^2 + (\eta b)^2 = 1 - 2\eta a + \eta^2 a^2 + \eta^2 b^2 \quad (A.5)$$

We require $|\mu|^2 < 1$, which implies:

$$\eta(a^2 + b^2) < 2a \quad (A.6)$$

and thus:

$$\eta < \frac{2a}{a^2 + b^2} = \frac{2\text{Re}(\lambda)}{|\lambda|^2} \quad (A.7)$$

This inequality must hold for every eigenvalue $\lambda$ of $A$. Therefore, to ensure convergence, we must choose $\eta$ such that:

$$0 < \eta < \min_{\lambda \in \sigma(A)} \frac{2\text{Re}(\lambda)}{|\lambda|^2} \quad (A.8)$$

where $\sigma(A)$ denotes the set of eigenvalues of $A$.

Since the set of eigenvalues is finite and each eigenvalue has a positive real part, the minimum is positive and thus such an $\eta > 0$ exists. With this choice of $\eta$, we have $|\mu| < 1$ for every eigenvalue $\mu$ of $G$, and hence $\rho(G) < 1$. Therefore, the iteration converges to the solution $\boldsymbol{q}_s$.

Similarly, the iteration scheme in Equation (12) also converges to $q_s$.